\documentclass[prd,twocolumn,
superscriptaddress,showpacs,nofootinbib, preprintnumbers]{revtex4}

\usepackage{amsmath}
\usepackage{amsfonts}
\usepackage{graphicx}
\usepackage{dcolumn}

\def\be{\begin{equation}}
\def\ee{\end{equation}}
\def\bea{\begin{eqnarray}}
\def\eea{\end{eqnarray}}
\def\bs{\begin{subequations}}
\def\es{\end{subequations}}

\def\nn{\nonumber \\}
\def\e{{\rm e}}

\begin{document}


\title{Phantom and non-phantom dark energy: The cosmological relevance of
non-locally corrected gravity}
\author{S. Jhingan}
 \affiliation{Centre for Theoretical
Physics, Jamia Millia Islamia, New Delhi, India}
\author{S. Nojiri}
 \affiliation{Department of physics, Nagoya university, Nagoya 464-8602, Japan}
\author{S.D. Odintsov}
 \affiliation{Instituci\`{o} Catalana de Recerca i Estudis Avan\c{c}ats (ICREA)
and Institut de Ciencies de l'Espai (IEEC-CSIC), Campus UAB,
Facultat de Ciencies, Torre C5-Par-2a pl, E-08193 Bellaterra
(Barcelona), Spain}
\author {M. Sami}
 \affiliation{Centre
for Theoretical Physics, Jamia Millia Islamia, New Delhi, India}
\author{I. Thongkool}
\affiliation{Centre for Theoretical Physics, Jamia Millia Islamia,
New Delhi, India}
\author{S. Zerbini}
\affiliation{Dipartimento di Fisica, Universit`a di Trento
and Istituto Nazionale di Fisica Nucleare
Gruppo Collegato di Trento, Italia}

\begin{abstract}
In this paper we have investigated the cosmological dynamics of
non-locally corrected gravity involving a function of the inverse
d'Alembertian of the Ricci scalar, $f(\Box^{-1} R))$. Casting the
dynamical equations into local form, we derive the fixed points of
the dynamics and demonstrate the existence and stability of a one
parameter family of dark energy solutions for a simple choice,
$f(\Box^{-1} R)\sim \exp(\alpha \Box^{-1} R)$. The effective EoS
parameter is given by, $w_{\rm eff}=({\alpha-1})/({3\alpha-1})$ and
the stability of the solutions is guaranteed provided that
$1/3<\alpha<2/3$. For $1/3<\alpha<1/2$ and $1/2<\alpha<2/3$, the
underlying system exhibits phantom and non-phantom behavior
respectively; the de Sitter solution corresponds to $\alpha=1/2$.
For a wide range of initial conditions, the system mimics dust like
behavior before reaching the stable fixed point. The late time
phantom phase is achieved without involving negative kinetic energy
fields. A brief discussion on the entropy of de Sitter space in
non-local model is included.
\end{abstract}
 \maketitle

\section{Introduction}
At present, there are two major theoretical approaches to late time
acceleration of universe which is supported by different data sets
of observations. The standard lore is related to the modification of
right hand side of the Einstein equations supplementing the stress
energy tensor by a {\it dark energy} component\cite{DE}. Recently,
serious attempts have been made to modify the geometry itself or the
original Einstein-Hilbert action. A large number of papers are
currently devoted to the investigations of $f(R)$ gravities(see
review\cite{review} and references therein). These theories are
motivated by phenomenological considerations. The problems faced by
these theories can be circumvented in specific models\cite{hu}. It
is really interesting that these models do not reduce to {\it
cosmological constant} $\Lambda$ in the low curvature regime and
thus can be distinguished from the latter. However, in any proposal
on modification of gravity at large scales, it becomes mandatory to
check whether the local gravity constraints are satisfied. The
latter can be achieved for specific $f(R)$ gravity models provided
one invokes the chameleon scenario\cite{hu,CT} {\it a la} Greek
epicycle.

It is also of crucial importance to explore the possibility of
obtaining late time acceleration from a fundamental theory. Thus the
string curvature corrections to gravity and their cosmological
relevance is a subject of current interest. Attempts have recently
been made to derive current acceleration using the Gauss-Bonnet (GB)
term and higher order curvature invariants coupled to a dynamically
evolving scalar field \cite{GB}. The model with GB invariant
exhibits a remarkable property that it does not disturb the scaling
regime and can give rise to late time transition from matter regime
to late time acceleration ({see Ref.\cite{TS}and references
therein). Unfortunately, the coupling of GB term to scalar field
gets large at late times; the nucleosynthesis also imposes stringent
constraints on these models. Inclusion of higher order corrections
introduces further technical complications \cite{SNO}.

Most of the proposals aimed to describe the late time cosmic
evolution are faced with one or the other problem. Recently, an
interesting idea of using non-locally corrected Einstein theory was
put forward in \cite{nonl,NO}. These corrections typically involve
combinations of inverse of d'Alembertian of the Ricci scalar and
might be induced by quantum loops and (or) stringy considerations.
Being non-local in character, these extra terms added to
Einstein-Hilbert action can lead to late time acceleration as a time
delayed effect avoiding the fine tuning problem\cite{nonl}. The
non-local dynamics can be cast in a local form by introducing a
number of auxiliary fields\cite{NO}.

In this paper, we consider the simplest form of non-local
corrections to gravity and investigate the underlying cosmological
dynamics in details. We explore the possibility of a matter like
regime which can finally mimic (phantom) dark energy and in
particular the de Sitter solution, at late times. The entropy of de
Sitter space in non-local gravity is also briefly discussed.

\section{Non-local cosmology and its late time attractors}
In what follows we consider the local form of non-locally corrected
gravity. Let us consider the following simple example of the
non-local action
 \cite{nonl}
\begin{equation}
\label{nl1}
S=\int d^4 x \sqrt{-g}\left\{
\frac{R}{2\kappa^2}\left(1+f(\Box^{-1}R)\right)+{\cal L}_{\rm matter}\right\}
\end{equation}
where $f$ is some function of d'Alembertian and  denoted by $\Box$.
The action would lead to non-local equations of motion which are
difficult to investigate. The above action can be cast into local
form  by introducing two scalar fields $\phi$ and $\xi$\cite{NO}
\begin{eqnarray}
\label{nl2}
S&=& \int d^4 x \sqrt{-g}\left[ \frac{1}{2\kappa^2}\left\{R\left(1
+ f(\phi)\right) + \xi\left(\Box\phi - R\right) \right\} \right. \nn
&& \qquad \qquad \qquad \qquad \qquad \qquad + {\cal L}_{\rm matter} \Bigr] \nn
&=& \int d^4 x \sqrt{-g}\left[
\frac{1}{2\kappa^2}\left\{R\left(1 + f(\phi)\right)
 - \partial_\mu \xi \partial^\mu \phi - \xi R \right\} \right. \nn
&& \qquad \qquad \qquad \qquad \qquad \qquad + {\cal L}_{\rm matter} \Bigr]\ .
\end{eqnarray}
The equivalence of actions (\ref{nl1}) and (\ref{nl2}) can easily be
checked. Indeed, by varying the action over $\xi$, we obtain
\begin{equation}
\label{boxphieq}
 \Box\phi=R~~ \textrm{or}~~ \phi=\Box^{-1}R,
\end{equation}
 which after substitution into
(\ref{nl2}), gives the original action (\ref{nl1}).

The local equations of motion can now be derived by varying with
respect to the metric and the auxiliary fields $\phi$ and $\xi$.
Variation of (\ref{nl2}) with respect to $g_{\mu\nu}$ gives
\begin{eqnarray}
\label{nl4}
0 &=& \frac{1}{2}g_{\mu\nu} \left\{R\left(1 + f(\phi) - \xi\right)
 - \partial_\rho \xi \partial^\rho \phi \right\} \nonumber \\
&& - R_{\mu\nu}\left(1 + f(\phi) - \xi\right)
+ \frac{1}{2}\left(\partial_\mu \xi \partial_\nu \phi +
\partial_\mu \phi \partial_\nu \xi \right) \nonumber \\
&& -\left(g_{\mu\nu}\Box - \nabla_\mu \nabla_\nu\right)\left(
f(\phi) - \xi\right) + \kappa^2T_{\mu\nu}\ .
\end{eqnarray}
The variation with respect to $\phi$ leads to
 \be \label{nl5}
0=\Box\xi+ f'(\phi) R\ . \ee For obvious reason, we shall specialize
to FRW background with the metric \be \label{nl6} ds^2 = - dt^2
+a(t)^2 \sum_{i=1,2,3}\left(dx^i\right)^2\ , \ee in which case the
scalar fields $\phi$ and $\xi$ are functions of time alone. In the
FRW background, the time-time and the space-space components of
Eq.(\ref{nl4}) yield the modified Hubble equation and equation for
acceleration
\begin{eqnarray}
\label{nl7a}
0 &=& - 3 H^2\left(1 + f(\phi) - \xi\right) - 3H\left(f'(\phi)\dot\phi - \dot\xi\right) \nn
&& + \frac{1}{2}\dot\xi \dot\phi + \kappa^2 \rho \ , \\
\label{nl7b}
0 &=& \left(2\dot H + 3H^2\right) \left(1 + f(\phi)
 - \xi\right) + \frac{1}{2}\dot\xi \dot\phi  \nonumber \nn
&& + \left(\frac{d^2}{dt^2} + 2H \frac{d}{dt} \right) \left( f(\phi)
 - \xi \right) + \kappa^2 p\ .
\end{eqnarray}
where $\rho$ and $p$ refer to the energy density and pressure of the
background fluid and satisfy the usual continuity equation
\begin{equation}
\label{eq:continuity}
\dot{\rho} + 3H(1 + w)\rho = 0
\end{equation}
with $w$ being the equation of state parameter of the background
matter. The scalar field equations assume the following form
\begin{eqnarray}
\label{nl8}
0 &=& \ddot \phi + 3H \dot \phi + 6 \dot H + 12 H^2 \ , \\
\label{nl8b} 0 &=& \ddot \xi + 3H \dot \xi - \left( 6 \dot H + 12
H^2\right)f'(\phi) \ .
\end{eqnarray}
For the sake of simplicity, we shall use the exponential form of
$f(\phi)$
\begin{equation}
\label{eq:falpha}
f(\phi) = f_0 \e^{\alpha \phi}\ ,
\end{equation}
where $f_0$ and $\alpha$ are constants. We shall now cast the
underlying dynamical equations in the autonomous form and look for
dark energy solution as stable fixed points of the dynamics. We use
the following notations (henceforth, we use the unit $\kappa^2=1$)
\begin{eqnarray}
N = \ln{a}\ ,\  x = \frac{\dot{\phi}}{H}\ ,\
y = \frac{\dot{\xi}}{H f(\phi)}\ ,\  z = \frac{\kappa^2 \rho}{H^2 f(\phi)}\ .
\end{eqnarray}
to arrive at the autonomous form of evolution equations
\begin{eqnarray}
\label{eq:dxn}
\frac{dx}{dN}&=& -3x - 12 - \frac{\dot{H}}{H^2}(x+6) \ , \\
\label{eq:dyn} \frac{dy}{dN}&=& -3y
+ 12\alpha-\alpha xy - \frac{\dot{H}}{H^2}(y - 6\alpha) \ , \\
\label{eq:dzn} \frac{dz}{dN}&=& -\left ( 3 + 3w +\alpha x +
2\frac{\dot{H}}{H^2} \right )z\ ,
\end{eqnarray}
and the constraint equation
\begin{equation}
\frac{\xi-1}{f(\phi)}=1 - \frac{1}{6}xy
+ \alpha x  - y - \frac{1}{3} z\ .
\end{equation}
Differentiating Eq.(\ref{nl7a}) with respect to time and using the
continuity equation (\ref{eq:continuity}), we can express
$\dot{H}/H^2$ in the term of the autonomous variables as
\begin{equation}
\label{eq:dh}
\frac{\dot{H}}{H^2} = \frac{ 24\alpha
+ 4\alpha x - \alpha^2 x^2 - (4+x)y -(1+w)z}
{2 \left [ \left ( 1 + \frac{x}{6} \right ) \left (
 -6\alpha + y \right ) + \frac{z}{3} \right ] }\ .
\end{equation}
First we shall analyze the dynamics in the absence of the background
fluid. In this case, the system (\ref{eq:dxn})$-$(\ref{eq:dyn})
possess two critical points,  $(x_c,y_c) = (0,0)$ and $(x_c,y_c) =
\left ( -2/\alpha , 6\alpha/(2-3\alpha) \right)$.
\begin{figure}[t]
\centering
\includegraphics[width=75mm]{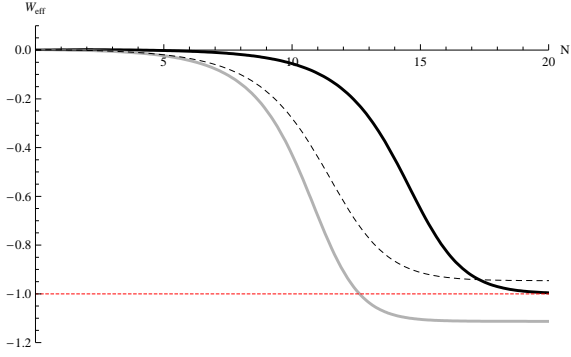}
\caption{The evolution of $w_{\rm eff}$ from the different initial
conditions, dark black line corresponds to $\alpha =0.5$, $x=-1.99$,
$y=5322.02$, gray to $\alpha =0.487$, $x=-1.99$, $y=1012.06$ and
dashed to $\alpha =0.507$, $x =-1.99$,$y=1013.06$. At the fixed
point $w_{\rm eff}$= $-1$, $-1.11$ and $-0.95$ respectively}
\label{fig:weffnoz}
\end{figure}
The stability of the critical points can easily be analyzed. The
eigenvalues of the stability matrix for the autonomous system
corresponding to the first critical point $(x_c,y_c) = (0,0)$ are
given by, $\lbrace -1,-1 \rbrace $ making it a  stable node. From
Eq.(\ref{eq:dh}), we easily find the effective EoS parameter at the
critical point under consideration
\begin{equation}
w_{\rm eff} = -1 - \left. \frac{2}{3}\frac{\dot{H}}{H^2}
\right|_{(0,0)} = \frac{1}{3}\ .
\end{equation}
For the fixed point $(x_c,y_c) = \left ( -2/\alpha ,
6\alpha/(2-3\alpha) \right)$, using Eq.(\ref{eq:dh}), we find
\begin{equation}
w_{\rm eff} = \frac{\alpha-1}{3\alpha-1}
\end{equation}
In this case, the eigenvalues of the stability matrix are given by,
$(3\alpha - 2)/(3\alpha - 1)$, $(3\alpha - 2)/(3\alpha - 1)$. Thus
the fixed point is stable provided that $1/3 < \alpha < 2/3$ leading
to dark energy solutions,  $-\infty < w_{\rm eff} < -1/3$; the
stable de Sitter solution is obtained for $\alpha = 1/2$, see
Fig.\ref{fig:weffnoz}. In Fig.\ref{fig:xyallnoz}, we have plotted
the phase portrait demonstrating the stability of both the critical
points. Without the loss of generality, we have assumed both $\phi$
and $\dot{\phi}$ to be negative which corresponds to a particular
type of boundary condition in Eq.(\ref{boxphieq})\cite{nonl}.
\begin{figure}[t]
\centering
\includegraphics[width=75mm]{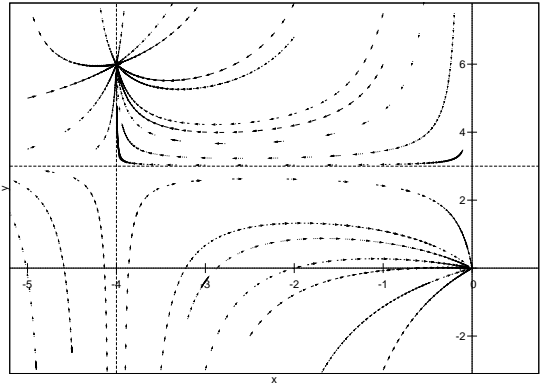}
\caption{The phase portrait of the system in absence of matter for
$\alpha=1/2$. The phase space splits into two disjoint regions: all
the trajectories starting from $y>3$ converge to de Sitter where as
they approach (0,0) for $y<3$ and $x>-4$. } \label{fig:xyallnoz}
\end{figure}
Inclusion of matter in the system makes the analysis cumbersome in
general. However, one can easily obtain the fixed point for $\dot{H}
= 0$
\begin{equation}
x_c = -4\ , \quad y_c = -\frac{12\alpha}{4\alpha - 3}\ , \quad z_c =
 -\frac{8\alpha(2\alpha - 1)}{1+w}\ .
\end{equation}
In this case, the eigenvalues of the stability matrix  are
\begin{equation}
\left \lbrace -3, 4\alpha - 3 , 4\alpha - 3 - 3w \right \rbrace\ .
\end{equation}
with a restriction that $\alpha = 3\left( 1 + w \right)/4$. In this
case, the fixed point is stable provided that $w<0$.

Despite the fact that it is extremely difficult to treat this system
analytically, several exact solutions may be found. For instance, in
case of the de Sitter solution, Eq.(\ref{nl8}) can be solved as \be
\label{NLdS1} \phi= - 4H_0t - \phi_0 \e^{-3H_0 t} + \phi_1\ , \ee
where $\phi_0$ and $\phi_1$ are the constants of integration. For
simplicity, we assume that $\phi_0=\phi_1=0$. Using Eq.(\ref{nl8b}),
we get the expression for $\xi$, \be \label{NLdS3} \xi= \frac{3f_0}{
4\alpha -3} \e^{-4\alpha H_0 t} - \frac{\xi_0}{3H_0}\e^{-3H_0 t} -
\xi_1\ . \ee Here $\xi_0$ and $\xi_1$ are constants. In the absence
of matter, $\rho=0$, the de Sitter space corresponds to $\alpha =
1/2$ and $\xi_1 = - 1$. When $\rho\neq 0$, there exists a de Sitter
solution provided we choose \be \label{NN2} \alpha =\frac{3}{4}(1+w)
\ ,\quad f_0 = \frac{\kappa^2 \rho_0}{3H_0^2 \left(1 + 3w \right)}\
,\quad \xi_1 = -1\ . \ee

In the presence of matter with $w\neq 0$, one may find a de Sitter
solution  even for a more complicated choice of $f(\phi)$
 \be
\label{nLL1} f(\phi)=f_0\e^{\phi/2} + f_1\e^{3(w+1)\phi/4}\ . \ee
The solution is given by
 \bea \label{nLL2} \phi &=& - 4H_0 t\ ,\quad \xi=1 +
3f_0\e^{-2H_0t}+ \frac{f_1}{w}\e^{-3(w+1)H_0 t}\ , \nn \rho &=& -
\frac{3(3w+1)H_0^2 f_1}{\kappa^2}\e^{-3(1+w)H_0t}\ . \eea

Let us now briefly point out some generalities of our formulation.
The above discussion of non-local cosmology is based upon the
simplest version of non-locality. Indeed, we could start from a
general non-local action: \bea \label{nl9} S &=& \int d^4 x
\sqrt{-g} \left[ F\left(R, \Box R, \Box^2 R , \cdots, \Box^m R,
\Box^{-1} R, \right. \right. \nn && \left. \left. \Box^{-2} R,
\cdots , \Box^{-n} R \right) + {\cal L}_{\rm matter} \right]\ . \eea
with $m$ and $n$ being the positive integers. By introducing scalar
fields $A$, $B$, $\chi_k$, $\eta_k$, $\left(k=1,2,\cdots,m \right)$,
and $\xi_l$, $\phi_l$, $\left(l=1,2,\cdots,n\right)$, we may rewrite
the action (\ref{nl9}) in the following form: \bea \label{nsigma1}
S&=&\int d^4 x \sqrt{-g} \left[ BR - BA + F\left(A, \eta_1, \eta_2,
\cdots, \eta_m, \right. \right. \nn && \left. \phi_1, \phi_2, \cdots
, \phi_n \right) + \partial_\mu \chi
\partial^\mu A + \sum_{k=2}^m
\partial_\mu \chi_k \partial^\mu \eta_{k-1} \nn && + \sum_{l=1}^n
\partial_\mu \xi_l \partial^\mu \phi_l + \sum_{k=1}^m \chi_1 \eta_1
+ A\xi_1 + \sum_{l=2}^n \xi_l \phi_{l-1} \nn && \left. + {\cal
L}_{\rm matter} \right] \ . \eea The variations over $A$, $B$
$\chi_k$, $\xi_l$ leads to the following equations
\bea
\label{nsigma2} && 0=R-A=\eta_1 - \Box A = \eta_k - \Box \eta_{k-1}
= \Box \phi_1 - A \nn && = \Box \phi_l - \phi_{l-1} \quad \left(k=2,
\cdots, m, \ l=2, \cdots, n\right)\ . \eea which gives $\eta_k =
\Box^k R$ and $\Box^l \phi_l = R$ and we find that the actions
(\ref{nsigma1}) and (\ref{nl9}) are equivalent.

Let us now consider a scale transformation given by \be
\label{nsigma3} g_{\mu\nu}\to \frac{1}{2 B} g_{\mu\nu}\ . \ee Then
the action (\ref{nsigma1}) is transformed into the action in the
Einstein frame: \bea \label{nsigma4} && S= \int d^4 x \sqrt{-g}
\left[ \frac{R}{2}
 - \frac{3}{2B^2}\partial_\mu B \partial^\mu B \right. \nn
&& + \frac{1}{2 B} \left(\partial_\mu \chi \partial^\mu A +
\sum_{k=2}^m \partial_\mu \chi_k \partial^\mu \eta_{k-1} +
\sum_{l=1}^n \partial_\mu \xi_l \partial^\mu \phi_l\right)  \nn && +
\frac{1}{4 B^2} \left( - BA + F\left(A, \eta_1, \cdots, \eta_m,
\phi_1, \cdots , \phi_n \right) \right. \nn && \left.\left. +
\sum_{k=1}^m \chi_1 \eta_1 + A\xi_1 + \sum_{l=2}^n \xi_l \phi_{l-1}
\right) + {\cal L}^A_{\rm matter} \right] \ . \eea Here ${\cal L}^A
_{\rm matter}$ can be obtained by scale transforming of the metric
tensor in ${\cal L}_{\rm matter}$ by (\ref{nsigma3}). The above
action can be regarded as a non-linear $\sigma$ model with potential
coupled to gravity. Action (\ref{nl1}) can be thought as a
particular case of the general non-local action. In this case,
however, the continuity equation (\ref{eq:continuity}) is no longer
valid; additional terms on the right hand side of this equation are
generated and are responsible for inducing interaction of matter
with background \cite{CT}. This aspect is crucial for the discussion
of local gravity constraints on the models based on the modified
theories of gravity.

\section{The entropy of  de Sitter space}

In this Section, we shall investigate the de Sitter space entropy
properties in non-local modified gravity with the cosmological
constant and matter.

The starting point is the trace of equation of motion of the
non-local model, which reads, for a constant curvature solution and
in the presence of a cosmological constant $\Lambda$ and matter, as
\bea
&& \left( f(\phi)+1 - \xi \right)R =4\Lambda -\kappa^2 T \nn
&& +6f'(\phi)R+ 3 f''(\phi)\partial_\rho \phi \partial^\rho \phi\ +
\partial_\rho \xi \partial^\rho \phi\,,
\label{t} \eea with $T$ the stress tensor trace. Let us suppose that
spherically-symmetric, static, constant curvature solution exists
\be ds^2=-A(r)dt_s^2+A(r)^{-1}dr^2+r^2d\Sigma^2 \,. \label{bh} \ee
There is a horizon if $A(r_H)=0$ with  $A'(r_H)\neq 0$ and $r_H
>0$. Then there is also an entropy associated with this horizon,
entropy that may be evaluated using Wald's method. For the local
action, one has \be {\cal S}_{BH}=\frac{A_H}{4G}\left[ 1-\xi+f(\phi)
\right]_{r_H}\,, \label{eds} \ee with all the fields being evaluated
on the horizon and $A_H=4\pi r_H^2$. Thus, in general, there is a
violation of the ``Area Law''.

Within the above black hole like metric, Eq. $\Box\phi=R$  and Eq.
for the field $\xi$ read for time dependent  and spherically
symmetric fields as \bea &&  -\frac{\partial_{t_s}^2
\phi}{A}+\frac{1}{r^2}
\partial_r \left(r^2 A \partial_r \phi\right)=R\,, \nn
&& -\frac{\partial_{t_s}^2 \xi}{A}+\frac{1}{r^2}
\partial_r \left(r^2 A \partial_r \xi\right)=-f'(\phi)R\,.
\label{b2}
\eea
One may look for the solution of the
first equation in the form $\phi=-4H_0 t_s-2\ln A(r)$. As a result,
it follows that   de Sitter space $A_{dS}(r)=1-Rr^2/12$ is a
solution. Then, it is easy to show that the
solution for $\xi$ reads
\be
\xi(r,t_s)=B+\frac{3f_0}{(4\alpha
 -3)}(1-H_0^2 r^2)^{-2 \alpha} \e^{-4 \alpha H_0 t_s}\,.
\ee As a result, the entropy factor becomes \be 1-\xi+f(\phi)=1-B+
\frac{(4 \alpha -6) f_0}{(4 \alpha -3)}(1-H_0^2 r^2)^{-2 \alpha}
\e^{-4 \alpha H_0t_s}\,. \label{s} \ee As a consequence, for
$\alpha>0$, the field $\xi$ and the entropy are, in general,
divergent on the horizon, while, as soon as $\alpha <0$, the $\xi $
field and the de Sitter entropy are finite and independent on the
time $t_s$ on the horizon. However, there exists a positive value,
namely $\alpha =3/2$, which renders the de Sitter entropy finite. We
will see the physical significance of these two  choices later.

Furthermore, in order to satisfy Einstein equations, the choice made
for $f(\phi)$ has to satisfy the trace constraint $(\ref{t})$, where
the stress tensor of the matter and cosmological constant
contribution are present. A direct computation shows that the trace
constraint is satisfied as soon as $ 1-B=\Lambda/3 H_0^2$, and the
stress tensor trace is not zero and is given by $T=T_0 A_{dS}(r)^{-2
\alpha}\e^{-4H_0 \alpha t_s}$, where $T_0$ depends on $ \alpha $.
Again, this trace diverges on the horizon as soon as $\alpha >0$.
Note that in the case $\alpha = \frac{3}{2}$, the entropy of the de
Sitter solution reads ${\cal
S}_{dS}=\left(A_H/4G\right)\left(\Lambda/3 H_0^2\right)$. For
vanishing cosmological constant we have found a de Sitter solution
with finite entropy which  is vanishing! However, recall the Wald
method gives the entropy modulo a constant.

In the FRW space-time, the solutions we have obtained lead to
\be
1-\xi+f(\phi) =\frac{\Lambda}{3H_0^2} +
\frac{\kappa^2(1-w) \rho_0}{3H^2_0w(1+3w)}\e^{-3H_0(1+w)t} \,,
\label{w}
\ee
while equation of motion for $\xi$ gives
\be
\label{nLL3}
1-\xi+f(\phi)=1-B+ f_0 \frac{4 \alpha -6}{4
\alpha-3}\e^{-4 \alpha H_0 t}\,.
\ee
A comparison with $w \neq 1$
gives
\be
1-B=\frac{\Lambda}{3H_0^2}\,, \ 4 \alpha =3(1+w)\,,
\ f_0=-\frac{k^2\rho_0}{3H_0^2(1+3w)}\,,
\ee
while if $w=1$, we
have $f_0$ undetermined. As a check,  we may pass to the static de
Sitter patch by transforming the time coordinate $t=t_s
+ \left(1/2 H_0\right)\ln (1-H_0^2r^2)$, and note
that one gets equation (\ref{s}) again.

Thus, the  existence conditions found for the de Sitter entropy
require a non vanishing positive cosmological constant and matter
and, in general, this matter, in order to have finite entropy,
should be ``phantom''matter, since the condition $\alpha <0$ is
equivalent to $1+w<0$. However, as already noted in the static
patch, if we have matter such  as $w=1$, namely stiff matter
$p=\rho$, important in the early universe, the time dependence of
the entropy drops out, in agreement with the result obtained  in the
static gauge.  One can also see that entropy divergences in the
static gauge correspond to the (non-physical) time-dependence of
entropy in the cosmological gauge and thus this situation seems
problematic. This consideration indicates that perhaps even  the
definition of entropy for non-local gravity should be reconsidered.
It is also interesting that AdS black hole solution for the theory
under consideration does not exist, while de Sitter solution exists.

\section{Conclusions}
In this paper we have investigated the cosmological relevance of
non-local corrections to gravity of the type $f(\Box^{-1} R)\sim
\exp(\alpha \Box^{-1} R)$ which can be motivated by non-perturbative
quantum effects. Casting the equations in local form we analyzed the
underlying dynamics and explored the possibility of finding stable
dark energy solutions. We have shown that the system might mimic the
matter dominated regime for a long time and ultimately switches over
to dark energy solutions at late times thereby alleviating the fine
tuning problem. The class of the dark energy solutions is
parameterized by $\alpha$ which ranges as $1/3< \alpha<2/3$. The
corresponding effective EoS parameter varies as, $-\infty <w_{\rm
eff}<-1/3$. In case of $\alpha>1/2$, we have non-phantom dark
energy; the de Sitter solution is obtained for $\alpha=1/2$. It is
remarkable that there exists a range of the parameter, namely,
$1/3<\alpha<1/2$ for which the model leads to phantom dark energy
solutions which are attractors of the system. It should be
emphasized that the observed acceleration can be achieved for
natural values of $\alpha \sim 1$. Secondly, the model under
consideration does not involve negative kinetic energy fields.
Unlike standard phantom theories giving rise to transient phantom
phase, the non-locally corrected gravity leads to a class of phantom
energy solutions which are the late time attractors of the dynamics.

It is also important to note that even if more precise observational
data defines the EoS parameter to be slightly different from $-1$,
there exists a possibility of realizing such scenario a in non-local
gravity as the effective quintessence or phantom cosmology.

Interestingly, the existence of de Sitter entropy in the non-local
model requires a non vanishing positive cosmological constant and
phantom matter.

Last but not the least, we should remember that the low energy
modifications of gravity crucially differ from the corresponding
local versions. In general, any large scale modification of gravity
faces two major challenges: The first is related to the presence of
either the ghost or the tachyon modes in these models; secondly, the
solar physics imposes stringent constraints on these models. Thus it
would be interesting to carry out these investigations for non-local
theories, as discussed in this paper; we defer this analysis to our
future work. It would also be interesting to investigate the
complicated versions of non-local gravity mentioned at the end of
section II.

\section*{Acknowledgments}
We thank V. Sahni and S. Tsujikawa for useful comments. This work is
supported by Japanese-Indian collaboration project (Grant No.
DST/INT/JSPS/Project-35/2007). The research by S.N. has been
supported in part by the Monbu-Kagaku-sho of Japan under grant
no.18549001 and 21st Century COE Program of Nagoya University
(15COEG01). The research by S.D.O. has been supported in part by the
projects FIS2006-02842 and PIE 2007-50/023(MEC, Spain). I.T.
acknowledges support under ICCR fellowship program.


\begin{thebibliography}{99}
\bibitem{DE} V.~Sahni and A.~A.~Starobinsky, Int.\ J.\ Mod.\ Phys.\ D
\textbf{9}, 373 (2000); S.~M.~Carroll, Living Rev.\ Rel.\ {}
\textbf{4}, 1 (2001); T.~Padmanabhan, Phys.\ Rept.\ {} \textbf{380},
235 (2003); P.~J.~E.~Peebles and B.~Ratra, Rev.\ Mod.\ Phys.\ {}
\textbf{75}, 559 (2003); E.~J.~Copeland, M.~Sami and S.~Tsujikawa,
Int. J. Mod. Phys., {D15} 1753 (2006); L. Perivolaropoulos,
astro-ph/0601014; J. Frieman, M. Turner and D. Huterer,
arXiv:0803.0982.
\bibitem{review} S.~Nojiri and S.~D.~Odintsov, Int.\ J.\ Geom.\ Meth.\ Mod.\
Phys.\ {\bf 4}, 115 (2007); arXiv:0801.4843[astro-ph].
\bibitem{hu} W. Hu and
I. Sawicki, Phys. \ Rev. \ D {\bf 76}, 064004 (2007).
\bibitem{CT} S. Capozziello and Shinji
Tsujikawa, arXiv:0712.2268.
\bibitem{GB}
S.~Nojiri {\it et al.}, Phys.\ Rev.\ D {\bf 71} (2005) 123509.
\bibitem{TS}  S. Tsujikawa,
hep-th/0606040.
\bibitem{SNO} M. Sami {\it et al.}, Phys. Lett. B {\bf 619} (2005); S.~Nojiri {\it et al.}, Phys.\ Rev.\ D {\bf 74}
046004 (2006); G.~Cognola {\it et al.}, Phys.\ Rev.\ D {\bf 75}
086002 (2007); E.~Elizalde {\it et al.}, Eur.\ Phys.\ J.\  C {\bf
53}, 447 (2008).
\bibitem{nonl}
S.~Deser and R.~Woodard, Phys.\ Rev.\ Lett.\ {\bf 99}, 111301 (2007)
arXiv:0706.2151.
\bibitem{NO} Shin'ichi Nojiri and S. D. Odintsov,
arXiv:0708.0924.



\end{thebibliography}
\end{document}